\documentclass[letterpaper, 10 pt, conference]{ieeeconf}  

\IEEEoverridecommandlockouts                              

\usepackage{graphicx} 
\usepackage{amsmath, bm} 
\usepackage{amsfonts}
\usepackage{amssymb}  
\usepackage{hyperref, soul}
\usepackage{pgfplots}
\usepackage{subcaption}
\pgfplotsset{compat=newest}
\title{\LARGE \bf
Intelligent Control of Differential Drive Robots Subject to Unmodeled Dynamics with EKF-based State Estimation}

\author{Amos Alwala$^{1}$, Yuchen Hu$^{1}$, Gabriel da Silva Lima$^{1}$, and Wallace Moreira Bessa$^{1}$
\thanks{“This project has received funding from the European Union’s Horizon Europe research and innovation programme under the Marie Skłodowska-Curie Actions grant agreement No. 101125250."}
\thanks{$^{1}$With the \href{https://sites.utu.fi/smartsystems/}{Smart Systems}, Department of Mechanical and Materials Engineering, Faculty of Technology,
        University of Turku, 20520 Turku, Finland.
        {\tt\small (amos.alwala, yuchen.y.hu, gdasil, wallace.moreirabessa)@utu.fi}}%
}

\begin{document}

\maketitle
\thispagestyle{empty}
\pagestyle{empty}

\begin{abstract}

 Reliable control and state estimation of differential drive robots (DDR) operating in dynamic and uncertain environments remains a challenge, particularly when system dynamics are partially unknown and sensor measurements are prone to degradation. This work introduces a unified control and state estimation framework that combines a Lyapunov-based nonlinear controller and Adaptive Neural Networks (ANN) with Extended Kalman Filter (EKF)-based multi-sensor fusion. The proposed controller leverages the universal approximation property of neural networks to model unknown nonlinearities in real time. An online adaptation scheme updates the weights of the radial basis function (RBF), the architecture chosen for the ANN. The learned dynamics are integrated into a feedback linearization (FBL) control law, for which theoretical guarantees of closed-loop stability and asymptotic convergence in a trajectory-tracking task are established through a Lyapunov-like stability analysis. To ensure robust state estimation, the EKF fuses inertial measurement unit (IMU) and odometry from monocular, 2D-LiDAR and wheel encoders. The fused state estimate drives the intelligent controller, ensuring consistent performance even under drift, wheel slip, sensor noise and failure. Gazebo simulations and real-world experiments are done using DDR, demonstrating the effectiveness of the approach in terms of improved velocity tracking performance with reduction in linear and angular velocity errors up to $53.91\%$ and $29.0\%$ in comparison to the baseline FBL.

\end{abstract}

\section{INTRODUCTION} \label{introduction}

Differential drive robots (DDR) have been widely used in warehouse logistics, autonomous navigation, research and education platforms owing to their mechanical simplicity and maneuverability~\cite{Schwab2021}. Their motion is however subjected to nonholonomic constraints; that is pure rolling without slip and no-lateral movement. These systems are also regarded as underactuated, as the number of control inputs is less than the required degrees of freedom (DOF)~\cite{alves_parametric_2018}. Based on these factors, design of robust control schemes for such DDR is complicated especially when the full system dynamics are partially or completely unknown, and when operating in dynamic and uncertain environments. 

\begin{figure}[!t!p]
\centerline{\includegraphics[width=1.8in]{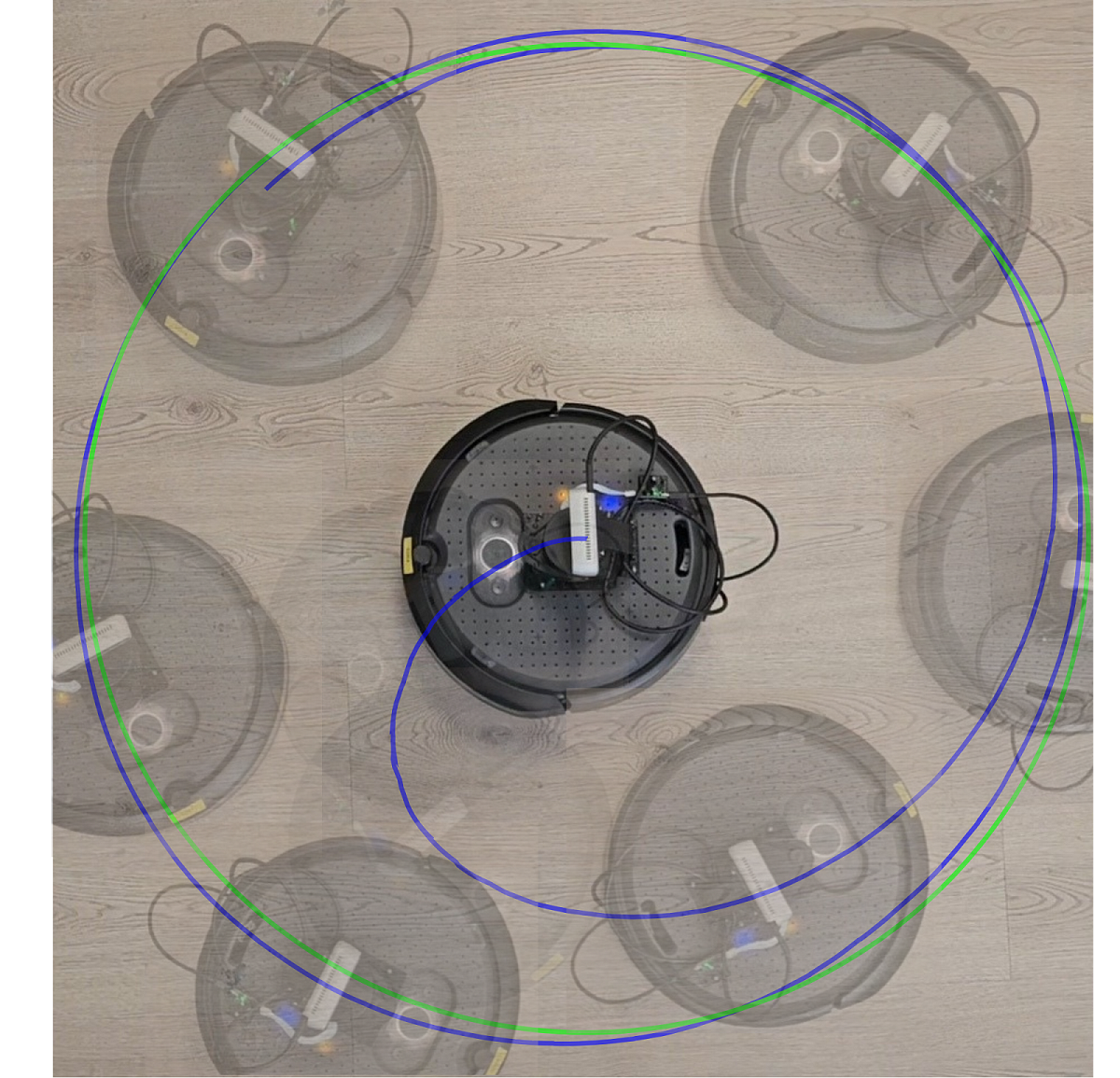}}
\caption{A time-lapse of the intelligent controlled Create3 DDR completing a circular trajectory}
\label{create3}
\end{figure}

Control approaches in the literature include kinematic control~\cite{Khatib2015, kanayama_stable_1990, Giuseppe2002}, fuzzy control~\cite{Zaman2020}, and sliding mode control (SMC)~\cite{doria-cerezo_sliding_2019}. 
In~\cite{doria-cerezo_sliding_2019}, a nonlinear SMC for path following was proposed using the orthogonal distance to the robot's longitudinal axis. The approach did not require robot global positioning and enabled bi-directional trajectory tracking, however, assumed perfect velocity tracking and employed a linearized observer. 
Other control approaches considered incorporating motor dynamics or applying dynamic linearization. For instance, ~\cite{park_discrete-time_2017} developed a discrete-time grey-box model with motor dynamics and a closed-form Coulomb friction term, assuming no slip, uniform surface and unicycle kinematics.
Likewise, ~\cite{alves_parametric_2018} applied recursive least-squares-parametrization on a linearized dynamic model.
An extension of Gauss' principle of least constraint was proposed in~\cite{zhang_computationally_2023},  incorporating holonomic and nonholonomic equality constraints within a nonlinear feedback control framework. The approach achieved high computational efficiency under high-speed maneuvers and disturbances but relied on a linear Karush-Kuhn-Tucker formulation for control synthesis.

With advances in neural networks (NN) and its applications to closed-loop control, data-driven and learning-based controllers have gained attention for capturing nonlinearities without explicit modeling. Typical implementations often employ offline training, where sensory inputs are mapped to control commands for behaviours such as obstacle avoidance. 
Foundation works such as~\cite{Fierro1998} introduced a Lyapunov-based combined kinematic/torque controller with multi-layer neural-network backstepping for DDR, and~\cite{Tamoghna2006} introduced a neuron-based adaptive controller with consideration of actuator dynamics. 
A class of NN, that is, Radial basis function (RBF) networks possess excellent universal approximation properties, making their use ideal for modeling and compensating unknown nonlinearities. 
In~\cite{da_silvalima_accurate_2023}, an RBF-based adaptive NN controller was proposed for an ominidirectional mobile robot trajectory tracking in presence of uncertainities.

Despite extensive progress in DDR control, most existing approaches rely on simplifying assumptions such as ideal velocity tracking, kinematics, linearized dynamics, or complete knowledge of the dynamics including actuation which limit robustness under real-world conditions involving model uncertainties, friction, and slippage. Learning-based and hybrid control strategies offer a promising alternative by leveraging data-driven adaptation while preserving model interpretability and stability guarantees.

In addition, a challenge lies in state estimation where relying on a single sensor leaves the robotic system vulnerable to sensor noise, drift, or sensor failure. Different sensing modalities present distinct advantages and limitations; fusing them enables the system to exploit complementary strengths while mitigating individual weaknesses. For instance, wheel encoders operate on the principle of measuring incremental wheel rotations to infer linear and angular displacement~\cite{Tran2022}. Their usage is light weight, computationally inexpensive and highly accurate over short durations. They however, accumulate error due to long-term drift, wheel slippage and any unmodeled nonholonomic effects while operating in uneven terrain. IMUs provide high-frequency measurements of linear accelerations and angular velocities which are valuable for short-term motion tracking. Nonetheless, they are prone to bias and disturbances from magnetic field interference leading to unbounded error if not filtered~\cite{Khatib2015}. Cameras/vision-based sensors extract environment features making them effective for localization and mapping, yet they degrade significantly in low-light and textureless environments~\cite{loquercio_learning_2021}. In contrast, LiDAR sensors operate by emitting laser pulses and measuring the time-of-flight to estimate distance from surrounding surfaces. This makes them robust to illumination changes though they can be computationally expensive and susceptible to degradation in adverse weather such as rain or fog.

Sensor fusion enables the control system to combine these complementary modalities thereby improving the robustness, consistency, and accuracy of state estimation. A number of fusion algorithms have been applied in literature including the Kalman Filter (KF), Unscented Kalman Filter, Extended Kalman Filter (EKF), and Particle Filter. Among these, the EKF is most widely adopted in robot localization application~\cite{menegatti_generalized_2016}, for its ability to handle nonlinear dynamics with relatively low computational burden.
Works such as~\cite{Khatib2015} demonstrate the effectiveness of EKF-based fusion of wheel encoder, compass, GPS and IMU. In~\cite{Tran2022} EKF was used for the fusion of wheel encoder readings and pose from markers detection by a camera.

In this paper, a hybrid framework that integrates online machine learning control with an EKF-based multi-sensor fusion, enabling both reliable control and state estimation in unstructured environments is presented. The Gaussian RBF neural network leverages the universal approximation property of RBF to model unknown nonlinearities including friction, unmodeled robot dynamics, actuator dynamics, and all other external disturbances in real-time. It employs an online adaptation scheme to update the weights of the RBF function. The controller utilizes a Lyapunov-based approach, which provides theoretical guarantees of closed-loop stability and asymptotic convergence in a trajectory-tracking task. The developed intelligent control algorithm is derived based on a first-order velocity dynamics model representing the velocity profile tracking. 
To ensure robust state estimation, the control law utilizes fused robot states provided by an EKF-based state estimator. The EKF ensures robust state estimation through a weighted fusion of IMU and odometry from wheel encoders, 2D-LiDAR and monocular sensing modalities, ensuring consistency even under drift, wheel slip, sensor noise, and failure. 

The presented approach is evaluated using the Robot Operating System (ROS)-2 with Gazebo simulations and real-world tests on DDRs. Despite incomplete knowledge of prior dynamics and the presence of external disturbances such as drift and unmodeled friction (smooth, rugged and soft floor), the proposed approach achieved precise trajectory tracking with reduced errors up to $80.67\%$ in $x$-position, $24.19\%$ in $y$-position, $82.48\%$ in yaw orientation, $53.91\%$ in linear velocity and $29.0\%$ in angular velocity, compared to the the pure feedback linearization.
The results discussed highlight the potential of combining adaptive machine learning control with state estimation to advance the reliability of autonomous robots.

The rest of the paper entails a discussion of an EKF-based state-estimation in Section~\ref{state estimation}. Robot kinematics and dynamics are presented in Section~\ref{differential drive robots}. In Section~\ref{control},  a Lyapunov-based control law is derived. Section~\ref{experiments and results} discusses the simulation and real experimental setups. Conclusions and future work are presented in Section~\ref{conclusions}.

\section{STATE ESTIMATION} \label{state estimation}
Accurate state estimation is crucial for robot control. This section describes an EKF framework employed to fuse the heterogeneous sensing modalities of IMU, wheel encoders, 2D-LiDAR, and monocular camera.

\subsection{EKF Formulation} 

The EKF is formulated based on the \textit{robot\_localization} ROS package~\cite{menegatti_generalized_2016} with the objective of estimating the 3-DOF pose and velocity of the DDR. The process is defined as a nonlinear dynamic system:
\begin{equation}\label{eq:ekf_model}
    \mathbf{x}_k = f(\mathbf{x}_{k-1}) + \mathbf{w}_{k-1}
\end{equation}
where $\mathbf{x}_k$ represents the 2D robot state vector at time $k$, $f$ represents a nonlinear state transition function and $\mathbf{w}_{k-1}$ represents the process noise which is assumed to be normally distributed. Measurement updates are received as:
\begin{equation}\label{eq:ekf_meas_up}
    \mathbf{z}_k = h(\mathbf{x}_k) + \mathbf{v}_k
\end{equation}
where $\mathbf{z}_k$ represents the measurement at time $k$, $h$ represents a nonlinear sensor model that maps the state into measurement space, and $\mathbf{v}_k$ represents the measurement noise which is assumed to be normally distributed.

Firstly, the prediction step projects the current state estimate and error covariance forward in time as:
\begin{equation}\label{eq:state_est_proj}
    \mathbf{\hat{x}}_k = f(\mathbf{x}_{k-1})
\end{equation}
\begin{equation}\label{eq:e_cov_proj}
    \bm{\hat{P}}_k = \bm{FP}_{k-1}\bm{F}^\top + \bm{Q}
\end{equation}
with $\bm{F}$, the Jacobian of $f$, used to project the estimate error covariance $\bm{P}$, and an addition of process noise covariance~$\bm{Q}$.
Then a correction step involving the computation of the Kalman gain $\bm{K}$ is defined as: 
\begin{equation}\label{eq:kalman_gain}
    \bm{K} = \bm{\hat{P}}_k\bm{H}^\top(\bm{H\hat{P}}_k\bm{H}^\top + \bm{R} )^{-1}
\end{equation}
where $\bm{H}$, the observation matrix, is the Jacobian matrix of the observation model function $h$, $\bm{R}$ is the measurement covariance, and $\bm{\hat{P}}$ is the projected error covariance.
This gain is then used to update the state vector and covariance matrix as: 
\begin{equation}\label{eq:state_vec_up}
    \mathbf{x}_k = \mathbf{\hat{x}}_k + \bm{K}(\bm{z}_k - \bm{H}\mathbf{\hat{x}}_k)
\end{equation}
\begin{equation}\label{eq:cov_mat_up}
    \bm{P}_k = (\bm{I} - \bm{KH})\bm{\hat{P}}_k(\bm{I} - \bm{KH})^\top + \bm{KRK}^\top
\end{equation}
The employed covariance update promotes stability by ensuring that $\bm{P}_k$ remains positive semi-definite.

\section{DIFFERENTIAL DRIVE ROBOTS} \label{differential drive robots}

\subsection{Robot Kinematics} \label{kinematics}

\begin{figure}[!t!p]
\centerline{\includegraphics[width=1.8in]{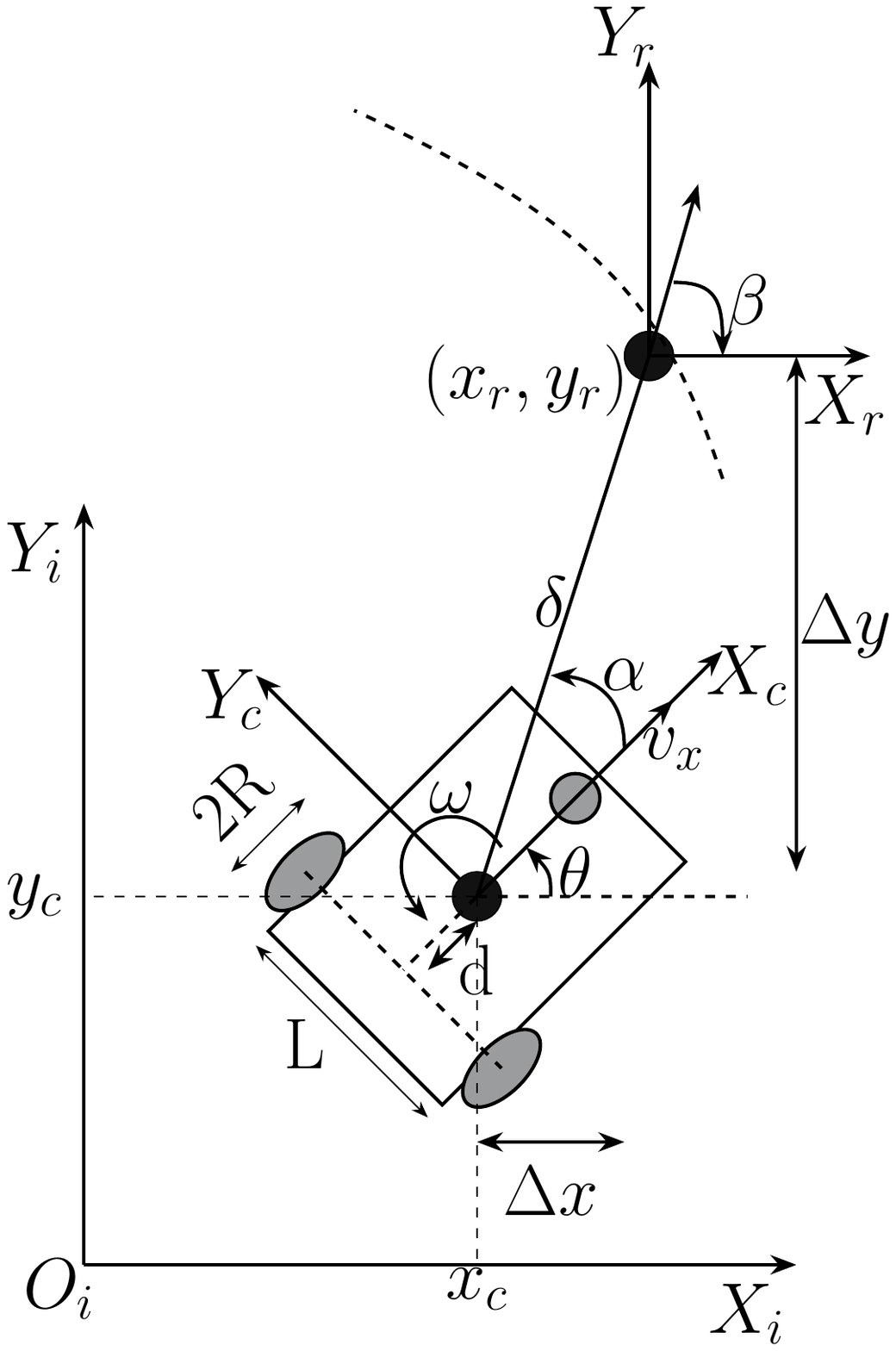}}
\caption{Schematic diagram of a differential drive robot}
\label{DDR} 
\end{figure}

Figure~\ref{DDR} shows the schematic representation of the differential drive robot with the coordinate system. It consists of two driving wheels of radius $R$ and a caster wheel. The motion and orientation of the DDR are achieved by independently actuated wheels whose separation is $L$ and mounted on the same axis with reference to the robot body frame.

The states of the robot are defined by a vector $\bm{q} = \begin{bmatrix}x_c &y_c &\theta\end{bmatrix}^\top$, where $(x_c, y_c)$ defines the position of the center of mass (COM) to the origin of the inertial plane $(X_i,Y_i)$, and $\theta$ defines the angle between the robot's axis of symmetry and the positive $X_i$-axis. The robot COM lies along $X_c$ at a distance $d$ from the midpoint $(x,y)$ of the axis connecting the wheels.

The nonholomonic behavior states that the robot can only move in the direction normal to the axis of the driving wheels i.e, the DDR satisfies the condition of pure rolling and no-slip. The kinematic constraints can be represented as 
\begin{equation}\label{eq:non-hol}
    \dot{y}_ccos\theta - \dot{x}_csin\theta - d\dot{\theta} = 0
\end{equation}
\begin{equation}\label{eq:non-hol_mat}
    \bm{A}(\bm{q})\bm{\dot{q}} = 0
\end{equation}

Given $\bm{S}(\bm{q})\in \mathbb{R}^{3\times2}$ a full rank matrix formed as a set of smooth and linearly independent vector fields spanning the null space of $\bm{A}(\bm{q})$,
\begin{equation}\label{eq:non-hol_mat_elim}
    \bm{S}^\top(\bm{q})\bm{A}^\top(\bm{q}) = 0
\end{equation}
from equations \eqref{eq:non-hol_mat} and \eqref{eq:non-hol_mat_elim}, it can be observed that an auxiliary vector time function $\bm{v}(t)$ exists such that
\begin{equation}\label{eq:qdot}
    \bm{\dot{q}} = \bm{S}(\bm{q})\bm{v}(t)
\end{equation}
where $\bm{v}(t) = \begin{bmatrix} v_x & \omega \end{bmatrix}^\top$ is the robot velocity vector which contains the linear and angular velocity responsible for motion control. $\bm{S(q)}$ is the Jacobian matrix that transforms the velocities from the mobile robot coordinates $\bm{v}$ to Cartesian coordinates $\bm{\dot{q}}$.

The kinematics equation of COM in terms of linear and angular velocity is given by
\begin{equation}\label{eq:Kinematics}
    \begin{bmatrix} \dot{x}_c \\\dot{y}_c \\\dot{\theta} \end{bmatrix} = \begin{bmatrix} cos\theta & -dsin\theta\\sin\theta & dcos\theta\\0 & 1 \end{bmatrix} \begin{bmatrix} v_x\\ \omega \end{bmatrix}
\end{equation}

\subsection{Robot Dynamics} \label{dynamics}

The dynamics model of the DDR is derived using the Lagrangian, $\frac{d}{dt}(\frac{\partial \mathcal{L}}{\partial \bm{\dot{q}}}) - \frac{\partial \mathcal{L}}{\partial \bm{q}} = \bm{F} - \bm{A}^\top\gamma$, where $\mathcal{L} = K_e - P_e$ is the Lagrangian, in which $K_e$ and $P_e$ stand for the kinetic and potential energies, respectively. For the robot operating in a horizontal plane, $P_e=0$ and the gravity term $\bm{G}(\bm{q})=0$. The kinetic energy is derived considering only the robot body of mass $m$, and neglecting the wheels and other dynamics.
\begin{equation}\label{eq:Ke}
    K_e = \frac{1}{2}mv^2 + \frac{1}{2}I_z\omega^2 =\frac{1}{2}m(\dot{x}_c^2 + \dot{y}_c^2) + \frac{1}{2}I_z\omega^2
\end{equation}
where $x_c = x + dcos\theta$ and $y_c = y + dsin\theta$.

By completing the Lagragian computation, the dynamics equation of the DDR can then be expressed in the form
\begin{equation}\label{eq:dynamics}
    \bm{M}(\bm{q})\bm{\ddot{q}} + \bm{C}(\bm{q, \dot{q}})\bm{\dot{q}} + \bm{\tau_d} = \bm{B}(\bm{q})\bm{\tau} - \bm{A}^\top(\bm{q})\bm{\gamma}
\end{equation}
where $\bm{M}(\bm{q}) \in \mathbb{R}^{3\times3}$ is a symmetric positive definite inertia matrix, $\bm{C}(\bm{q,\dot{q}}) \in \mathbb{R}^{3\times3}$ is the centripetal and coriolisis matrix, $\bm{\tau_d} \in \mathbb{R}^{3\times1}$ represents friction, external disturbances, and other unmodeled dynamics including actuators and wheels, $\bm{B}(\bm{q}) \in \mathbb{R}^{3\times2}$ is the input transformation matrix, $\bm{\tau}$ is the input vector, $\bm{A}(\bm{q}) \in \mathbb{R}^{3\times1}$ is the matrix associated with the nonholonomic constrains, and $\bm{\gamma}$ is the vector of constrained forces.

The dynamics \eqref{eq:dynamics} is transformed into a first-order system suitable for control by differentiating \eqref{eq:qdot} and substituting for $\bm{\ddot{q}}, \bm{\dot{q}}$.
Further multiplication by $\bm{S}^\top$ eliminates the nonholonomic constraint matrix $\bm{A}^\top\bm{\gamma}$.
The DDR system in a new set of local coordinates becomes:
\begin{equation}\label{eq:model}
    \bm{\bar{M}}(\bm{q})\bm{\dot{v}} + \bm{\bar{C}}(\bm{q,\dot{q}})\bm{v} + \bm{\bar{\tau}_d} = \bm{\bar{B}}(\bm{q})\bm{\tau}
\end{equation}
$$\label{eq:inertial_mat}
    \bm{\bar{M}}(\bm{q})=\begin{bmatrix}m &0 \\0 &I_z+md^2\end{bmatrix}, \bm{\bar{C}}(\bm{q, \dot{q}})=\begin{bmatrix}0 &-2md\dot{\theta} \\2md\dot{\theta} &0\end{bmatrix},$$
$$\label{eq:inertial_matr}
    \bm{\bar{B}}(\bm{q}) = \frac{1}{R}\begin{bmatrix}1 &1 \\L &-L\end{bmatrix}$$

\section{ROBOT CONTROL} \label{control}

\subsection{Intelligent Control} \label{intelligent control}
Intelligent controllers have the ability to learn by interacting with their environment, thereby predicting the outcome of their actions. These have found application in robot control for trajectory tracking~\cite{da_silvalima_accurate_2023}, actuator position tracking~\cite{dos_santos_intelligent_2019}, depth control of diving agents~\cite{bessa_design_2017}, among other uncertain nonlinear systems. 

For the model presented in the \eqref{eq:model}, the feedback linearization control law\cite{slotine1991applied} is proposed below:
\begin{equation}\label{eq:int_ctl_law}
    \bm{\tau} = \bm{\bar{B}}^{-1}\bm{\bar{C}v} + \bm{\bar{B}}^{-1}\bm{\bar{M}}(\bm{\dot{v}_r} - \bm{\Lambda\tilde{v}} - \bm{\hat{d}})
\end{equation}
where $\bm{\dot{v}}_r$ is the derivative of the reference velocity, $\bm{\tilde{v}}=\bm{v}-\bm{v}_r$ is the velocity tracking error, $\bm{\Lambda} \in \mathbb{R}^{2\times2}$ is a diagonal matrix with positive entries $\lambda_i$, $\bm{d}$ represents all disturbances including friction and unmodeled dynamics and $\bm{\hat{d}}$ is the estimate of $\bm{d}$.

By applying the control law \eqref{eq:int_ctl_law} to robot dynamics \eqref{eq:model}, the error dynamics becomes:
\begin{equation}\label{eq:error_dyn}
    \bm{\dot{\tilde{v}}} + \bm{\Lambda\tilde{v}} = \bm{d} - \bm{\hat{d}}
\end{equation}
It can be verified that as $\bm{\hat{d}}\to \bm{d}$ in an ideal estimation, the tracking error $\bm{\tilde{v}}$ exponentially converges to zero. If not, closed-loop dynamics is driven by the approximation error $\bm{\hat{d}} - \bm{d}$.

To turn the control law into an intelligent controller, we introduce the use of an RBF network universal approximator to estimate all uncertainty $\bm{d}$. 
\begin{equation}\label{eq:dhat}
    \hat{d}_i = \mathbf{w}^\top_i\bm{\varphi}_i(v_i)
\end{equation}
where $\hat{d}_i$ are the components of $\bm{\hat{d}}$, $ \mathbf{w}_i = \begin{bmatrix}w_{i,1} &w_{i,2} & \dots &w_{i,n_i}\end{bmatrix}$ is the weight vector and $\bm{\varphi}_i(v_i)=\begin{bmatrix}\varphi_{i,1} &\varphi_{i,2} &\dots &\varphi_{i,n_i}\end{bmatrix}$ is the vector with the activation functions $\varphi_{i,j} = \exp[-\frac{(v_i-c_{i,j})^2}{2\sigma^2_{i,j}}]$, with $i = 1, 2$, $j = 1, 2, ..., n_i$ and $n_i$ being the number of neurons in the hidden layer corresponding to $i$th component of $\bm{\hat{d}}$.

We assume an ideal set of weights $\mathbf{w}^*$ that allows for an exact approximation of $\bm{d} = \mathbf{w}^{*\top}\bm{\varphi}$. Therefore, the estimation error becomes $\tilde{\bm{d}}= \hat{\bm{d}} - \bm{d} = (\mathbf{w} - \mathbf{w}^*)^\top\bm{\varphi}= \tilde{\mathbf{w}}^\top\bm{\varphi}$, which shows that the estimation error depends on how far the current weights are from the ideal.

A Lyapunov function is proposed to prove the boundedness and convergence conditions of the closed-loop system as:
\begin{equation}\label{eq:lyapunov_can}
    V_i = \frac{1}{2}\tilde{v}^2_i + \frac{1}{2\eta_i}\tilde{\mathbf{w}}^\top_i\tilde{\mathbf{w}}_i
\end{equation}
whose time derivative becomes:
\begin{equation}\label{eq:der_lyapunov}
    \dot{V}_i = \tilde{v}_i\dot{\tilde{v}}_i + \eta^{-1}_i\mathbf{\tilde{w}}^\top_i\mathbf{\dot{w}}_i
\end{equation} since $\mathbf{w}^*$ is constant. After further respective substitutions, we obtain:
\begin{equation}\label{eq:der_lyapunov1}
    \dot{V}_i = \tilde{v}_i(-\lambda_i{\tilde{v}_i} - \tilde{d}_i) + \eta^{-1}_i\mathbf{\tilde{w}}^\top_i\mathbf{\dot{w}}_i
\end{equation}
\begin{equation}\label{eq:der_lyapunov2}
    \dot{V}_i = \tilde{v}_i(-\lambda_i{\tilde{v}_i} - \tilde{\mathbf{w}}^\top_i\bm{\varphi}_i) + \eta^{-1}_i\mathbf{\tilde{w}}^\top_i\mathbf{\dot{w}}_i
\end{equation}
\begin{equation}\label{eq:der_lyapunov3}
    \dot{V}_i = -\lambda_i\tilde{v}^2_i + \eta^{-1}_i\mathbf{\tilde{w}}^\top_i(\mathbf{\dot{w}}_i - \eta_i\tilde{v}_i\bm{\varphi}_i)
\end{equation}
Therefore, by updating $\mathbf{w}_i$ according to the learning rule $\mathbf{\dot{w}}_i = \eta\bm{\tilde{v}}\bm{\varphi}_i$, with $\eta$ as the learning rate. The time derivative $\dot{V}_i$ becomes:
\begin{equation}\label{eq:der_lyapunov4}
    \dot{V}_i = -\lambda_i\tilde{v}^2_i
\end{equation}
implying that $\dot{V}_i$ is negative definite and continues to decrease as $\bm{\tilde{v}} \to 0$ and $t \to \infty$.

\section{EXPERIMENTAL SETUP AND RESULTS} \label{experiments and results}
The performance of the proposed control scheme is firstly evaluated in simulation and later on hardware. The simulations were done using ROS2 Humble with Gazebo Ignition and ROS2 Jazzy with Gazebo Harmonic using the \href{https://github.com/AntoBrandi/Bumper-Bot}{bumperbot} model. Real-world indoor experiments were performed using ROS2 Humble on Jetson Orin Nano 8GB with the iRobot Create3 mounted with 2D RPLidar-A8 and Realsense D435 camera.

\subsection{Trajectory Tracking} \label{trajectory}
We consider a circle trajectory tracking problem defined by $\begin{bmatrix}c_x+rcos\omega t &c_y+rsin\omega t &\theta_r\end{bmatrix}$, where $(c_x,c_y)$ is the center of the circular trajectory, $r$ is the radius of the circle, $\omega$ is the angular velocity and $\theta_r$ is the orientation defined as $\tan^{-1}\left(\frac{\dot{y}}{\dot{x}}\right)$.
The trajectory tracking task as shown in Fig.~\ref{DDR} is to find a smooth velocity control such that $\lim_{t \to \infty}(\bm{q_r} - \bm{q}) = 0$. Then compute $\bm{\tau}$, the torque input required  so that $\bm{v} \to \bm{v}_c$ as $t \to \infty$.

\subsection{Simulation} \label{simulation}
\begin{figure}[!t!p]
\centerline{\includegraphics[width=2.9in]{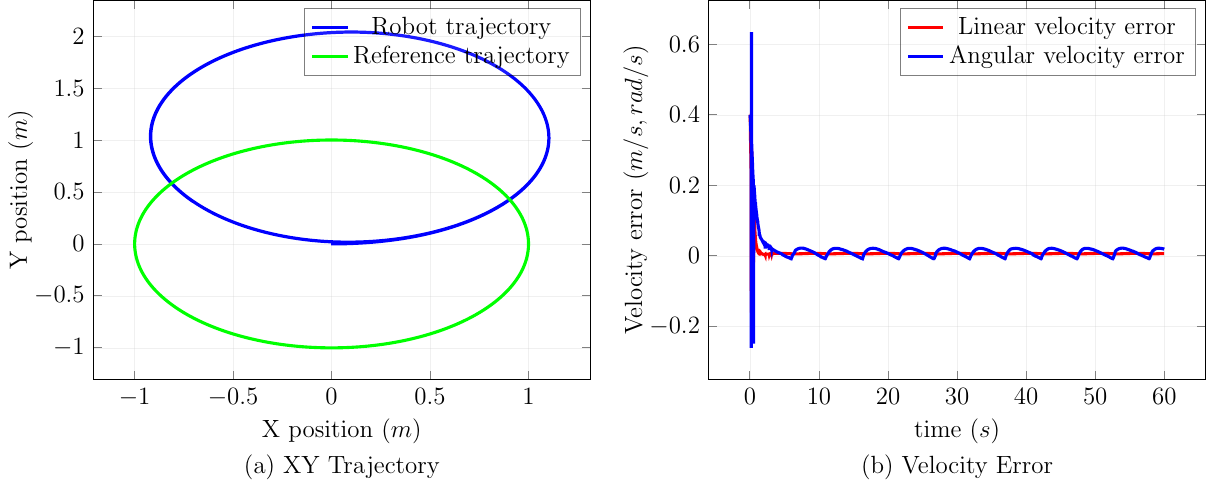}}
\caption{Gazebo simulation results showing trajectory tracking and velocity error of the intelligent controller.}
\label{before_vel_ctl}
\end{figure}

\begin{figure}[!t!p]
\centerline{\includegraphics[width=2.9in]{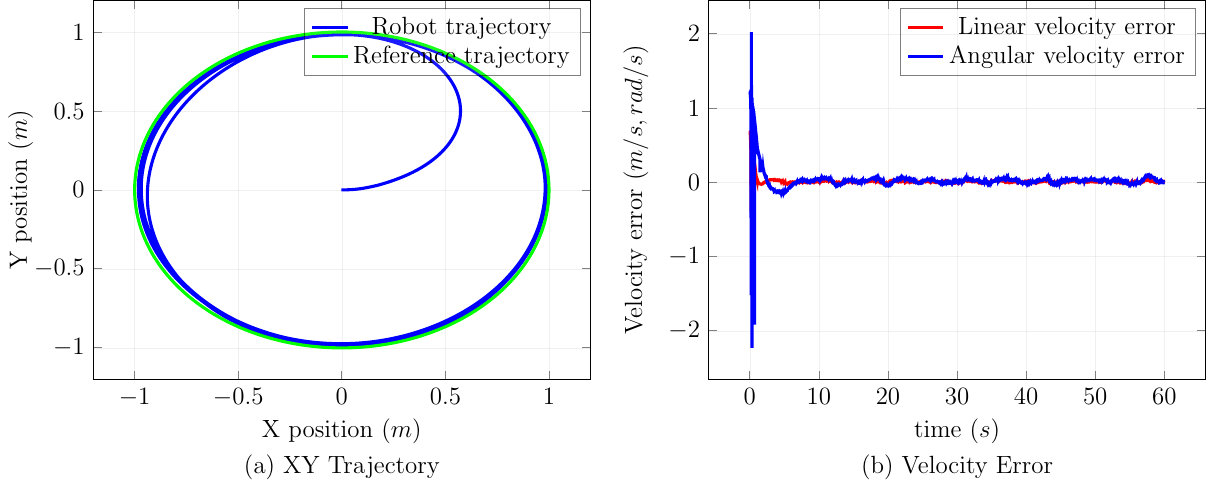}}
\caption{Gazebo simulation results showing trajectory tracking and velocity error of the intelligent controller with an intermediate pose feedback velocity controller.}
\label{int_vel_ctl}
\end{figure}

Gazebo physics simulation and the Unified Robot Description Format (URDF) were used to model the robot's dynamics and environmental interactions such as inertia, joint constraints, and surface friction. The robot states were obtained by an EKF-based fusion of IMU and wheel odometry. DDR parameters were set as $m$ = 0.10054\,kg, $R$ = 0.034\,m, $L$ = 0.17\,m, $I_z$ = 0.003\,kg/m$^3$, whereas, the controller parameter is set as $\lambda_i = 3$. Two layers of the RBF network are considered, one for each state variable with a single input $v_i$, a single output $\hat{d}_i$, and six neurons in the hidden layer chosen to balance a trade-off between computation and control accuracy. The centers and widths of the Gaussian function were as $\bm{c}_i = \begin{bmatrix}-1.0 &-0.5 &-0.25 &0.25 &0.5 &1.0\end{bmatrix}$, $\bm{\sigma}_i = \begin{bmatrix}0.3 &0.2 &0.1 &0.1 &0.2 &0.3\end{bmatrix}$. The weight vector is initialized as $\mathbf{w}_i = 0$ to allow the robot learn from interaction with its environment with a learning rate $\eta_i = 10$. Fig.~\ref{before_vel_ctl} shows the tracking performance of the circle trajectory with $r$ = 1.0\,m and $\omega$ = 0.4\,rad/s. It can be seen that the controller does not track the reference trajectory because the control law \eqref{eq:int_ctl_law} does not consider direct pose feedback, however, the velocity error is seen to decay as expected.

To introduce pose feedback to the intelligent control scheme~\ref{intelligent control}, we employ an intermediate velocity control law\cite{kanayama_stable_1990} that computes $\bm{v_c}$ from the trajectory tracking error as expressed below:
\begin{equation}\label{eq:kanayama_vel}
    \bm{v}_c = \begin{bmatrix}v_dcose_\theta +k_1e_x \\\omega_d + k_2v_de_y + k_3v_dsine_\theta\end{bmatrix}
\end{equation}
\begin{equation}\label{eq:kanayama_e}
    \begin{bmatrix}e_x \\e_y \\e_\theta\end{bmatrix} = \begin{bmatrix}cos\theta &sin\theta & 0 \\-sin\theta &cos\theta &0 \\0 &0 &1\end{bmatrix}\begin{bmatrix}x_r - x_c \\y_r - y_c \\\theta_r - \theta\end{bmatrix}
\end{equation}
where $k_1, k_2, k_3 > 0$. 
The tracking error is computed from the difference between the reference trajectory pose and the current robot pose, transformed to a local coordinate system whose origin is $(x_c, y_c)$.
The resulting velocity is then used to compute the velocity error used in the intelligent control scheme, $\bm{\tilde{v}}=\bm{v_c}-\bm{v_r}$. The trajectory tracking results of the intelligent controller utilizing the intermediate velocity controller are shown in Fig.~\ref{int_vel_ctl} with gains as $k_1= 0.5, k_2= 1.0, k_3=1.5$.

\subsection{Real-world} \label{real world}
The performance of the intelligent controller is then evaluated experimentally on the iRobot Create3 DDR. The robot consists of an IMU, an optical mouse and wheel encoders for motion detection. The Create3 ROS2 API exposes raw sensor topics of which we consider the \texttt{odom}, a dead-reckoning estimate of pose computed on-board from fusion of the mentioned raw sensors. Additional sensors are mounted vertically above the base footprint of the DDR. The robot coordinate system corresponds to the right-hand with $x$ forward, $y$ left and $z$ up.

\subsubsection{EKF localization}
\href{https://github.com/AlexKaravaev/ros2_laser_scan_matcher}{Lidar odometry} (LO) uses the Point-to-Line Iterative Closest/Corresponding Point (PLICP) algorithm~\cite{Andrea_lidar_odom_2008}, whereas monocular visual odometry (VO) is based on \href{https://github.com/UZ-SLAMLab/ORB_SLAM3}{ORB-SLAM3}~\cite{campos_orb-slam3_2021} taking advantage of its multiple map system with an improved place recognition approach suitable for long periods of poor visual information. 
The EKF measurement update step reads the odometry topics and computes a change in state by integrating linear velocities for position update and angular velocities for orientation update.
A key feature of the \textit{robot\_localization ekf\_localization\_node} is that it permits partial state vector updates based on per input configuration of $\bm{H}$. 
In addition, the process noise covariance $\bm{Q}$ is exposed as a tunable parameter. 
An initial EKF estimate covariance $\bm{P}_{i,j} = 10^{-9}$ and the process noise covariance $\bm{Q}_{i,j} = 0.05$ are used. This allows the EKF to slowly respond to sensor measurements while trusting its motion model more. The covariances are maintained and updated at each time-step.
The EKF correction step then applies measurement updates from sensor readings. Wheel odometry is configured to update the velocities $(v_x, v_y, w)$, whereas, LO and VO absolute pose readings $(x,y,\theta)$ are configured to update velocities and not absolute pose to avoid any jumps that might occur during LO/VO.

The EKF weights these updates based on associated measurement covariances $\bm{R}$ from each sensor odometry topic and uncertainty $\bm{P}$. If the associated measurement covariance is small, the EKF considers the sensor reading to be accurate and leans more towards this, otherwise if the measurement covariance is large, the EKF smoothly blends this state measurement with other state updates.
This enables the EKF to handle estimation loss experienced by VO as a result of feature loss leading to map re-initialization, especially in low-texture or low light environment. Herein, the EKF is designed to degrade under partial sensor failure, ensuring that the remaining sensors compensate, therefore enabling continued state estimation. The fused state estimate is then utilised by the intelligent control scheme.

\subsubsection{Intelligent control}
\begin{figure}[!t!p]
\centerline{\includegraphics[width=2.9in]{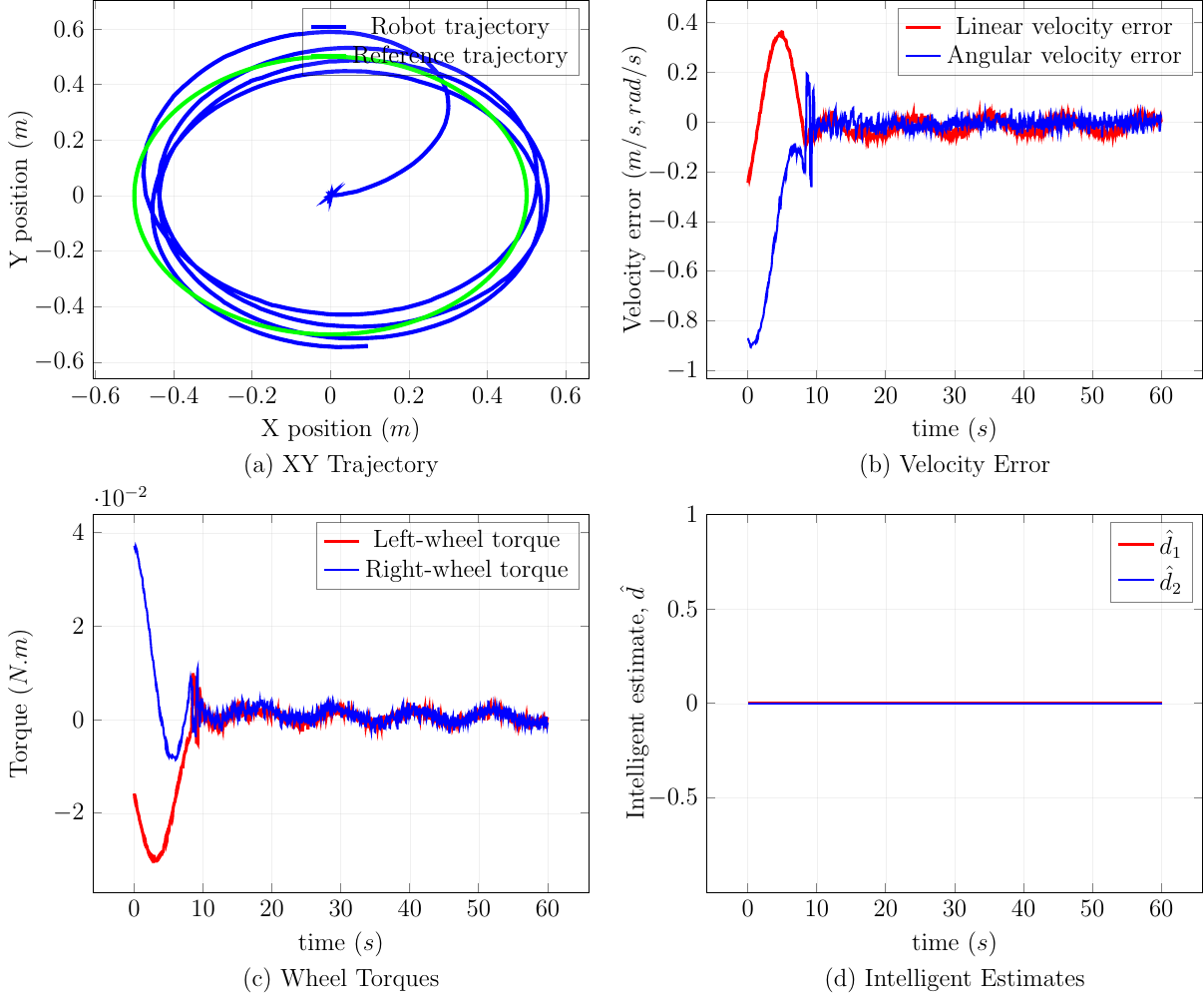}}
\caption{Results of feedback linearization controller on Create3 tracking a circle trajectory of $0.5 m$ radius at $0.5 rad/s$ angular speed.}
\label{C3_fbl} 
\end{figure}

\begin{figure}[!t!p]
\centerline{\includegraphics[width=2.9in]{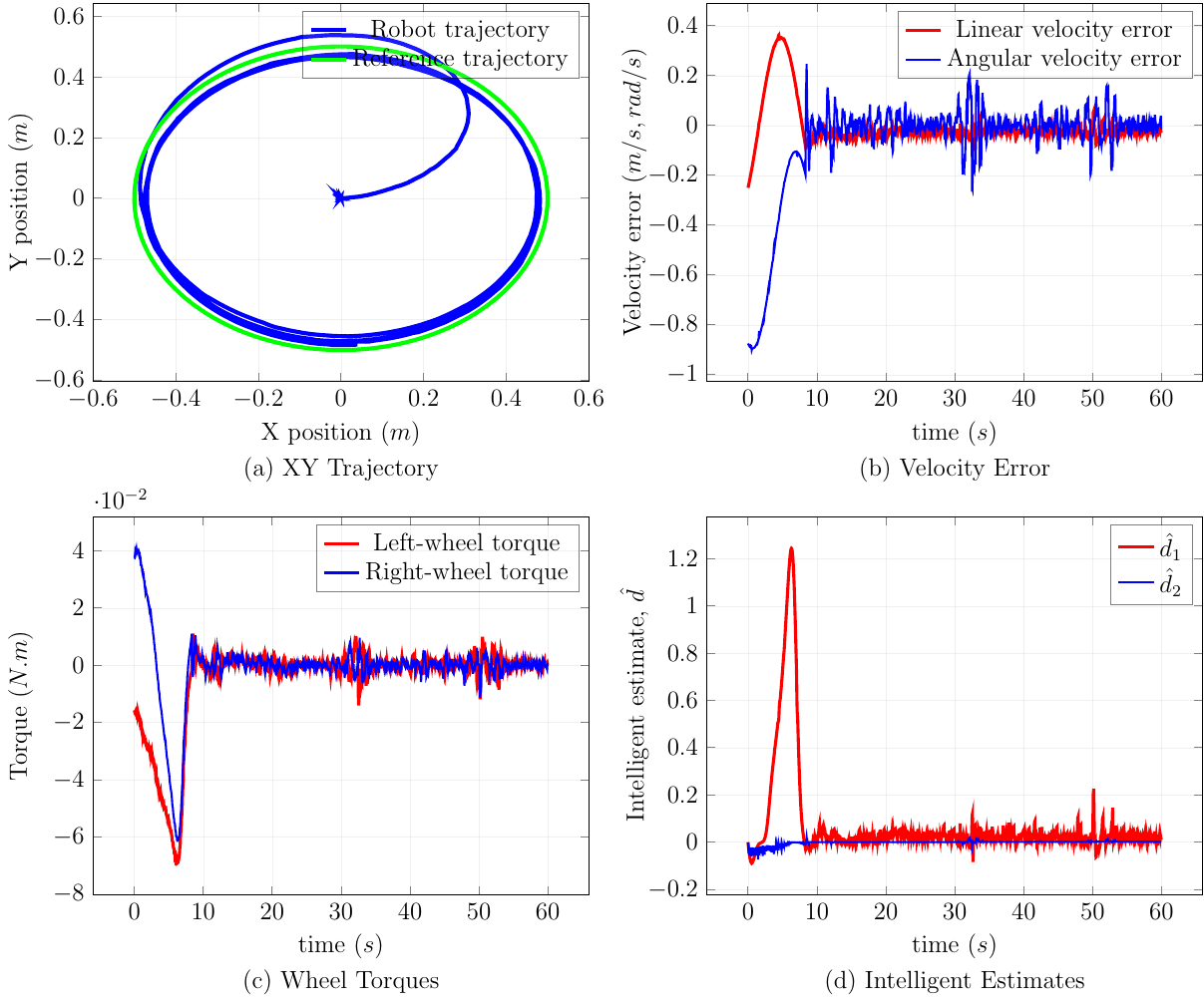}}
\caption{Results of the intelligent controller on Create3 tracking a circle trajectory of $0.5 m$ radius at $0.5 rad/s$ angular speed.}
\label{C3_ic}
\end{figure}

The robot parameters are set as $m$ = 2.5\,kg, $R$ = 0.035\,m, $L$ = 0.287\,m, $I_z$ = 0.5\,kg/m$^3$, and for the controller $\lambda_i = 1.0$, with the widths and centers as in simulation, learning rate $\eta_i = 0.8$ and no prior knowledge of friction and incomplete dynamics. Because Create3 expects control input through the \texttt{cmd\_vel} topic, a conversion is applied to obtain the linear and angular velocity expected by the robot. The computed velocities are saturated considering the linear \texttt{max\_speed} of 0.46\,m/s of the DDR configured at runtime from the web server. The reference trajectory is also defined to take into account this velocity constraint with $r$ = 0.5\,m and $\omega$ = 0.5\,rad/s. The control scheme is evaluated with increased velocity and different floor characterists (i.e. smooth, rugged, and soft). This variation introduces disturbance to the system causing challenges to the feedback linearization scheme, however, the intelligent control scheme deals with these variations as illustrated by the results.

The trajectory tracking results of the intelligent control scheme are shown in fig.~\ref{C3_ic}. The impact of the intelligent controller is evaluated by comparing with trajectory tracking results in fig.~\ref{C3_fbl} when $\eta_i = 0$, a pure feedback linearization controller. It can be seen that the intelligent controller is capable of tracking the desired trajectory with improved performance in terms of velocity tracking error. A comparison of the steady state root mean squared error in terms of position, orientation, linear, and angular velocity is shown in table~\ref{error_comp} for both the intelligent controller and the pure feedback linearization controller. From these results, the effect of the Gaussian RBF in being able to compensate for unmodeled disturbances is evident. The results also show that the fusion enabled controller maintains stability even though a sensor failure occurred as observed with VO re-initialization.

\begin{table}[h]
\caption{A comparative analysis of the controller trajectory tracking performance on rugged floor with $\eta_i = 0.8$ in terms of Steady State Root Mean Squared Error}
\label{error_comp}
\begin{center}
\begin{tabular}{c c c c }
\hline
 & Pure Feedback Linearization & Intelligent & Change (\%)\\
\hline
$E_x$ & 0.0712 & 0.0138 & 80.67\\
$E_y$ & 0.0416 & 0.0315 & 24.19\\
$E_\theta$ & 0.1242 & 0.0218 & 82.48\\
$E_v$ & 0.0416 & 0.0192 & 53.91\\
$E_\omega$ & 0.055 & 0.0391 & 29.00\\
\hline
\end{tabular}
\end{center}
\end{table}

\section{CONCLUSIONS} \label{conclusions}
This paper presents a unified control and estimation framework applicable to mobile robots operating under uncertain dynamics and unreliable sensing. An online RBF NN controller was developed to approximate unknown nonlinearities with a Lyapunov-based controller design that ensures closed-loop stability and asymptotic convergence of trajectory tracking errors. To complement the intelligent control scheme, an EKF that fuses IMU, monocular, 2DLiDAR and wheel odometry was employed, enabling robustness of the state estimation against drift and sensor failure. Simulation and real-world experiments on a DDR platform confirmed that the proposed approach achieves good trajectory tracking ability, error convergence, and resilience against unmodeled disturbances. Future works will include the use of reinforcement learning to tune controller parameters and integrate this framework with higher-level path planning applications such as is required in SLAM.


\section*{ACKNOWLEDGMENT} \label{acknowledgment}

“Funded by the European Union. Views and opinions expressed are however those of the author(s) only and do not necessarily reflect those of the European Union or European Research Executive Agency (REA). Neither the European Union nor the granting authority can be held responsible for them.”

"We acknowledge the financial support of the Finnish Ministry of Education and Culture through the Intelligent Work Machines Doctoral Education Pilot Program (IWM VN/3137/2024-OKM-4)."

\bibliographystyle{IEEEtran} 
\bibliography{IEEEabrv, ECC26_ref} 

\addtolength{\textheight}{-12cm}   

    
\end{document}